\documentclass[12pt]{article}

\usepackage[dvips]{color,graphicx}
\usepackage{psfrag}
\usepackage{amssymb}
\usepackage{axodraw}

\definecolor{gris}{rgb}{0.8,0.8,0.8}

\newcommand{\eq}[1]{\begin{equation} #1 \end{equation}}

\newcommand{\eqa}[1]{\begin{eqnarray} #1 \end{eqnarray}}

\newcommand{\av}[1]{\langle #1 \rangle}

\newcommand{\ds}{\displaystyle}

\newcommand{\heff}{\mathcal{H}_{\rm eff}}

\newcommand{\op}{\mathcal{O}}
\newcommand{\nn}{\nonumber}
\newcommand{\Li}{{\rm Li_2}}
\renewcommand{\sl}[1]{#1 \hspace{-0.2cm}/}
\newcommand{\mg}{m_{\tilde g}}
\newcommand{\ms}{\widetilde m}

\begin{document}


$\ $
\vspace{1.5cm}
\begin{center}
\Large\bf 
Top mass dependent ${\cal O}(\alpha_s^3)$ corrections to\\ $B$-meson mixing in the MSSM
\end{center}

\vspace{3mm}
\begin{center}
{\sc Javier Virto
}
\end{center}

\begin{center}
{\em 
Universitat Aut\`onoma de Barcelona, 08193 Bellaterra, Barcelona, Spain
}
\end{center}

\vspace{1mm}
\begin{abstract}\noindent

We compute the top mass dependent NLO strong interaction matching conditions to the $\Delta F=2$ effective Hamiltonian in the general MSSM. We study the relevance of such corrections, comparing its size with that of previously known NLO corrections in the limit $m_t\to 0$, in scenarios with degeneracy, alignment, and hierarchical squarks. We find that, while these corrections are generally small,  there are regions in the parameter space where the contributions to the Wilson coefficients $C_1$ and $C_4$ could partially overcome the expected suppression $m_t/M_{SUSY}$.

\end{abstract}


\section{Introduction}

The phenomenon of neutral meson mixing has been the subject of extensive studies within the field of flavor physics in the last decades, and has provided a good deal of knowledge concerning the flavor structure of the SM and of its extensions \cite{report}. Experimental data on the mixing of $K$ and $B_d$ mesons constrain very strongly the existence of  sources of flavor violation beyond Minimal Flavor Violation within the first to families, pushing the scale of generic flavor violating New Physics far from the TeV scale \cite{0707.0636,blum,Isidori:2010kg}.

On the other hand, the knowledge we have of flavor transitions between the second and the third generation of quarks is not so precise. Besides theoretical reasons to expect New Physics to show its face in these type of transitions, there are already several experimental tensions that could be the effect of physics beyond the SM (see for example \cite{ckmfit,ciuchini}). An important focus of the LHC era is to probe $B_s$ physics to a novel level of precision, with the objective of clarifying these issues. In this context, the progress towards higher theoretical precision in $B_s$ physics is necessary. 

In the case of the Minimal Supersymmetric Standard Model (MSSM) without model dependent assumptions for the mecanism of supersymmetry breaking, the  $\Delta F=2$ effective Hamiltonian  describing neutral meson mixing is known to NLO in strong interactions. Leading order (LO) strong interaction matching conditions in the MSSM have been known for some time, and were computed in Refs.~\cite{Gerard:1984bg,hep-ph/9604387,Hagelin:1992tc}; they arise from squark-gluino box diagrams. The corresponding $\Delta F=2$ next-to-leading order (NLO) corrections come from two loop diagrams, and have been computed in Ref.~\cite{hep-ph/0606197} within the Mass Insertion Approximation, and in the general case in Refs.~\cite{JV,Virto:2009iv}. The anomalous dimension matrix for the complete set of operators has been computed at NLO in QCD in Refs.~\cite{hep-ph/9711402,hep-ph/0005183}, and a resummation of large logarithms related to large squark mass ratios in scenarios with hierarchical squark masses can be found in Ref.~\cite{bertuzzo}. A taste of the flavor constraints on the SUSY parameter space in the general MSSM can be found, for example, in Refs.~\cite{report,Silvestrini:2007yf,Crivellin:2009ar,arXiv:0812.3610}.

The present time is a particularly appropriate moment to reanalyze the phenomenology of SUSY models because of the direct searches for SUSY particles that are being performed at the LHC. The experimental collaborations  ATLAS and CMS have already reported exclusion limits based on $\sim 1/{\rm fb}$ of data in searches for $jets + \sl{E}_T$  with zero \cite{atlas0l,cms0l} or more charged leptons, and larger datasets up to $\sim 15/{\rm fb}$ are expected in the future.

These limits, however, are model dependent, and the lower bounds on the SUSY masses vary from model to model. In the constrained MSSM (CMSSM), squarks and gluinos of equal mass are excluded below $\sim 950\,{\rm GeV}$ \cite{atlas0l}, but the bound on the gluino mass would be slightly higher ($\sim 1.1\,{\rm TeV}$) for squarks around $500\,{\rm GeV}$ \cite{cms0l}. A more detailed analysis including LHCb data, as well as electroweak precision and $B$-physics observables  obtains best fit values of $(m_{1/2},m_0)\sim(780,450)$ for the CMSSM and $(m_{1/2},m_0)\sim(730,150)$ for the NUHM1 \cite{isidoriLHC}. In the case of the phenomenological MSSM (pMSSM), detailed analyses find that $85\%$ of the parameter space is excluded for gluino masses below $\sim 520\,{\rm GeV}$ \cite{arbeyLHC}, while $m_{\tilde g,\tilde q}\gtrsim 1.1\,{\rm TeV}$ if $\mg\sim m_{\tilde q}$ and $\mg\gtrsim 700\,{\rm GeV}$ if $m_{\tilde q}\gg \mg$ \cite{pieriniLHC}. Thus, the SUSY scale can still be low enough for flavor physics to impose substantial constraints on flavor-violating soft supersymmetry-breaking couplings in general versions of the MSSM, and these constraints should certainly be reevaluated. At the same time this reevaluation should involve, as argued above, a novel level of theoretical precision. 

In Ref.~\cite{JV} the $\Delta F=2$ NLO strong interaction matching conditions were computed, putting all the quark masses to zero. This approximation is entirely justified since corrections to this limit are suppressed by a ratio $m_q/M_{SUSY}\ll 1$. In the case of the top quark this suppression still exists, although it might not be so effective if the SUSY scale is below the TeV, with $m_t/M_{SUSY}\sim 1/6$. In such a case, it may be that the loop functions have numerical enhancements that compensate such suppression: a factor of $\sim 10$ is not unreasonable, at least in some regions of the parameter space. The only way to know if this happens is to compute explicitly these corrections.

The purpose of this paper is to clarify this issue, by making the two loop calculation and comparing the size of the corrections with the $m_t\to 0$ limit. After reviewing briefly the structure of the $\Delta F=2$ effective Hamiltonian in Section \ref{sec:Heff}, we will describe the computation of the $m_t$-dependent corrections in Section \ref{sec:Det}, specifying the structure of the Wilson coefficients. The explicit expressions for the $m_t$-dependent Wilson coefficients in the general case are presented in Appendix \ref{sec:res}, and in the Mass Insertion Approximation in Appendix \ref{sec:resMIA}. In Section \ref{sec:size} we analyze the size of the corrections in three different SUSY scenarios: degeneracy, alignment, and a model with hierarchical sfermions. Conclusions and a summary of the results are presented in Section \ref{sec:con}.

\section{Effective Hamiltonian for $\Delta F=2$ processes}
\label{sec:Heff}

The most general effective Hamiltonian for  $\Delta F=2$ processes up to operators of dimension six can be written as
\eq{\heff^{\Delta F=2}=\sum_{i=1}^5 C_i\, \op_i+\sum_{i=1}^3\tilde{C}_i\,\tilde{\op}_i
\label{Heff}}
where $C_i$ are the Wilson coefficients and $\op_i$ are the dimension six $\Delta F=2$ operators. In four dimensions there are eight independent operators of this type. Here we choose the following basis:
\eq{
\begin{array}{rclrcl}
\op_1& = & \bar s_\alpha \gamma_\mu P_Lb_\alpha\ \bar s_\beta \gamma^\mu P_Lb_\beta \qquad &
\tilde\op_1 &=& \bar s_\alpha \gamma_\mu P_R b_\alpha\ \bar s_\beta \gamma^\mu P_R b_\beta\\[2mm]
\op_2& = & \bar s_\alpha P_Lb_\alpha\ \bar s_\beta P_Lb_\beta &
\tilde\op_2& = & \bar s_\alpha P_R b_\alpha\ \bar s_\beta P_R b_\beta\\[2mm]
\op_3& = & \bar s_\alpha P_Lb_\beta\ \bar s_\beta P_Lb_\alpha &
\tilde\op_3& = & \bar s_\alpha P_Rb_\beta\ \bar s_\beta P_Rb_\alpha\\[2mm]
\op_4& = & \bar s_\alpha P_Lb_\alpha\ \bar s_\beta P_Rb_\beta &
\op_5& = & \bar s_\alpha P_Lb_\beta\ \bar s_\beta P_Rb_\alpha 
\end{array}
\label{ops}
}
where $P_{L,R}=(1\mp\gamma_5)/2$ are the usual chiral projectors. To simplify the notation throughout the paper we focus on the case of $B_s-\bar B_s$ mixing, while other cases of interest can be recovered by obvious substitutions of quark fields.

The mixing amplitude is obtained by taking the matrix element of the effective Hamiltonian between $B$ and $\bar B$ states:
\eq{A=\sum_i C_i(\mu)\,\av{{\cal O}_i(\mu)}
=\sum_{ij} C_j(\Lambda)\,U(m_b,\Lambda)_{ij}\,\av{{\cal O}_i(m_b)}
\label{Amp}}
where $\Lambda$ is the matching scale at which the Wilson coefficients are computed. The matrix $U(m_b,\Lambda)$ is the evolution matrix, that accounts for the renormalization group running of the coefficients down to the hadronic scale. The evolution matrix at NLO can be found for example, in Ref.~\cite{Becirevic:2001jj}. The matrix elements of the operators can be found in Ref.~\cite{0707.0636}. These matrix elements are calculated in the RI-MOM scheme, which means that the Wilson coefficients used in Eq.~(\ref{Amp}) must be computed in the this scheme.

The NLO Wilson coefficients presented here and in Ref.~\cite{JV} are given in the NDR scheme. The prescription to translate between NDR, DRED and RI-MOM schemes involves rotating the set of coefficients with appropriate ${\cal O}(\alpha_s)$ matrices, which are given for example in Ref.~\cite{hep-ph/0606197}.

\section{Top mass dependent corrections}
\label{sec:Det}

\begin{figure}
\begin{center}
\psfrag{b}{$b$} \psfrag{bb}{$\bar b$}
\psfrag{s}{$s$} \psfrag{sb}{$\bar s$}
\psfrag{t}{$t$} \psfrag{k}{$\tilde q_k$}
\psfrag{g}{$\tilde g$} \psfrag{i}{$\tilde q_i$}
\psfrag{j}{$\tilde q_j$}
\includegraphics[width=6cm,height=3.5cm]{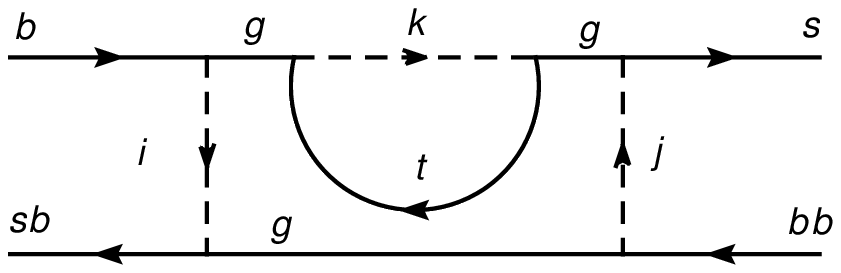}\hspace{1.8cm}
\includegraphics[width=6cm,height=3.5cm]{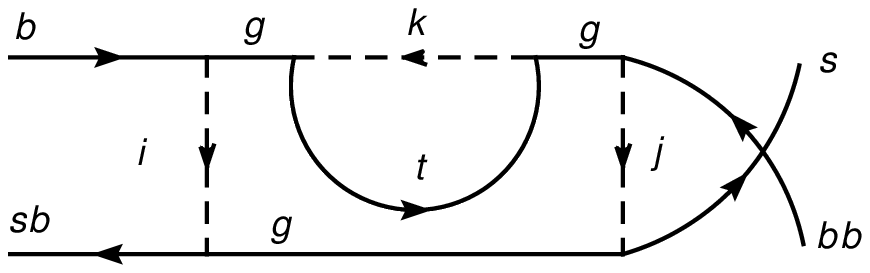}
\end{center}
\caption{A representative sample of the two loop diagrams that contribute to the ${\cal O}(\alpha_s^3)$ top mass dependent corrections to the mixing of $B_s$ mesons.}
\label{diags}
\end{figure}

At two loops, the top quark appears for the first time inside the Feynman diagrams that contribute to the $\Delta F=2$ effective Hamiltonian. It appears in the one loop correction to the gluino propagator, in box diagrams such as those depicted in Fig.~\ref{diags}.

A diagram of this kind is proportional to the following two loop integral:
\eq{{\cal D}\sim\int d^4q_1 d^4q_2\ \frac{[V^{js}(\sl{q}_1+\mg)U^{kt}(\sl{\Delta q}+m_t)V^{kt}
(\sl{q}_1+\mg)U^{ib}]
\otimes [V^{is}(\sl{q}_1+\mg)U^{jb}]}{(q_1^2-\mg^2)^3(q_1^2-\ms_i^2)(q_1^2-\ms_j^2)
(q_2^2-\ms_k^2)(\Delta q^2 - m_t^2)}}
where $\Delta q = q_1-q_2$, $U^{iq}=\Gamma_L^{iq}P_L-\Gamma_R^{iq}P_R$, and $V^{iq}=\Gamma_L^{iq*}P_R-\Gamma_R^{iq*}P_L$. The squark rotation matrices $\Gamma$ are defined as being the rotations that relate squark fields in the super-CKM basis $(\tilde q_{i,L}^I,\tilde q_{i,R}^I)$ to the mass eigenstates $\tilde q_i$:
\eqa{
\tilde d_{i,L}^I=\Gamma_{D_L}^{j i*}\tilde d_{j}\ ,&\quad& \tilde d_{i,R}^I=\Gamma_{D_R}^{j i*}\tilde d_{j}\nn\\
\tilde u_{i,L}^I=\Gamma_{U_L}^{j i*}\tilde u_{j}\ ,&\quad& \tilde u_{i,R}^I=\Gamma_{U_R}^{j i*}\tilde u_{j}
}
 The masses and momenta can be normalized to the gluino mass: defining $x_i \equiv \ms_i^2/\mg^2$ and $x_t\equiv m_t^2/\mg^2$, we have
\eq{
{\cal D}\sim\frac{1}{\mg^2}\int d^4q_1 d^4q_2\ \frac{[V^{js}(\sl{q}_1+1)U^{kt}
(\sl{\Delta q}+\sqrt{x_t})V^{kt}(\sl{q}_1+1)U^{ib}]
\otimes [V^{is}(\sl{q}_1+1)U^{jb}]}{(q_1^2-1)^3(q_1^2-x_i)(q_1^2-x_j)
(q_2^2-x_k)(\Delta q^2 - x_t)}
}
The leading corrections for $x_t\to 0$ where computed in Ref.~\cite{JV}. Expanding up to ${\cal O}(m_t^2/m_{\tilde g}^2)$ one gets
\eq{{\cal D}\sim {\cal D}_{x_t=0}+\frac{\sqrt{x_t}}{\mg^2}\int d^4q_{1,2}\ 
\frac{\cdots\ U^{kt} V^{kt}\ \cdots}{(\cdots)\Delta q^2}
+\frac{x_t}{\mg^2}\int d^4q_{1,2}\ \frac{\cdots\ U^{kt} \Delta\sl{q}\, V^{kt}\ \cdots}
{(\cdots)\Delta q^4} + \cdots
}
where we have kept the pieces from the squark-top loop.

We see that the $m_t/\mg$ contribution is proportional to $\Gamma^{kt*}_L\Gamma^{kt}_R$ and $\Gamma^{kt*}_R\Gamma^{kt}_L$, while the $m_t^2/\mg^2$ contribution is proportional to $\Gamma^{kt*}_L\Gamma^{kt}_L$ and $\Gamma^{kt*}_R\Gamma^{kt}_R$. This means that the $m_t/\mg$ correction is suppressed by a $LR$ mass insertion, as expected for the required chiral flip.

The corresponding contributions to the Wilson coefficients in Eq.~(\ref{Heff}) from all these diagrams can be written as
\eqa{
C_\ell^{(t)}=&&\frac{\alpha_s^3}{12^3 \pi \mg^2} \sum_{ijk}\Bigg[
a_\ell^{ijk}(\Gamma^{ijk}_{ABCDLL}+\Gamma^{ijk}_{ABCDRR})+\frac{m_t}{\mg} (g_\ell^{ijk} \Gamma^{ijk}_{ABCDLR} + h_\ell^{ijk} \Gamma^{ijk}_{ABCDLL})\nn\\
&&\hspace{1.6cm} + \frac{m_t^2}{\mg^2} f_\ell^{ijk} (\Gamma^{ijk}_{ABCDLL}+\Gamma^{ijk}_{ABCDRR}) + \cdots \Bigg]
\label{WCst}}
where $\Gamma^{ijk}_{ABCDEF} \equiv \Gamma_A^{is*} \Gamma_B^{ib}\, \Gamma_C^{js*} \Gamma_D^{jb}\, \Gamma_E^{kt*} \Gamma_F^{kt}$  (capital letters mean $L$ or $R$), and the sum runs over 6 down-type squarks in the case of $i,j$ and 6 up-type squarks for $k$.

In Eq.~(\ref{WCst}), the functions $a_\ell^{ijk}$ correspond to the $m_t=0$ contribution, and are given  explicitly in Ref.~\cite{JV}. The functions $f_\ell^{ijk}$, $g_\ell^{ijk}$ and $h_\ell^{ijk}$ have been calculated here, and are given explicitly in Appendix \ref{sec:res}. The details of this calculation are the same as those described in Ref.~\cite{JV}, and we refer the interested reader to that reference.

\section{Size of the corrections}
\label{sec:size}

In this section we will study the size of the $m_t$-dependent corrections. Naively, one expects these corrections to be at the few percent level with respect to the $m_t=0$ correction: the first term is of order $m_t/m_{\tilde g}\lesssim 20\%$, but contains a further suppression from the LR mass insertion; the second term is of order $m_t^2/m_{\tilde g}^2\lesssim {\rm few}\%$. Constraints on the $LR$ mass insertion come from vacuum stability considerations, which require that \cite{hep-ph/9606237}:
\eq{\delta^{tt}_{LR}\lesssim 2\frac{m_t}{\ms}\ ,}
but the exact bounds depend on the particular spectrum of the model. Therefore, the two contributions have roughly the same suppression. The question is whether the loop functions can account for a compensating enhancement, at least over some regions of the parameter space.

In order to address this issue, we will explore the SUSY parameter space in three different scenarios: degeneracy, alignment and hierarchy.

\subsection{Degeneracy}

We consider the Wilson coefficients in the Mass Insertion Approximation, where the squark squared mass matrix in the super-CKM basis has degenerate diagonal entries much bigger than the off diagonal terms (mass insertions). The Wilson coefficients depend on a reduced set of parameters: the average squark mass $M_s$, the gluino mass ---through the parameter $x\equiv \mg^2/M_s^2$, and the mass insertions $\delta^{sb}_{LL}$, $\delta^{sb}_{RR}$, $\delta^{sb}_{RL}$, $\delta^{sb}_{LR}$ and $\delta^{tt}_{LR}$. The Wilson coefficients in the MIA can be obtained from the general expressions presented in Appendix \ref{sec:res}. The general form of these coefficients is:
\eq{C_i^{\rm MIA} = \frac{\alpha_s^3}{12^3\pi M_s^2}
\Bigg[a_i(x)+\frac{m_t}{\mg}\ g_i(x)\ {\rm Re}(\delta^{tt}_{LR})
+\frac{m_t^2}{\mg^2}\ f_i(x)+\cdots \Bigg] \delta_{AB}^{sb}\delta_{CD}^{sb}\ ,}
where $A,B,C,D=L,R$. The functions $a_i(x)$ correspond to the $m_t=0$ contribution, and can be found in Ref.~\cite{hep-ph/0606197}.\footnote{In this reference the coefficients are given in the DRED scheme. One should convert them to the NDR scheme to be consistent with the conventions in the present paper.} The functions $f_i(x)$ and $g_i(x)$ can be found in Appendix \ref{sec:resMIA}, where the complete expressions for the Wilson coefficients in the MIA are given.

In order to estimate the size of the new corrections, we consider the following ratios, $R^g_i(x)\equiv g_i(x)/a_i(x)$ and $R^f_i(x)\equiv f_i(x)/a_i(x)$, and determine if there is a region in the variable $x$ where any of these ratios are large.

\begin{figure}[t!]
\begin{center}
\psfrag{x}{$x$} \psfrag{r1}{\small \hspace{-0.5cm} $R_1^g$, $R_1^f$}
\psfrag{r2}{\small \hspace{-0.5cm} $R_2^g$, $R_2^f$}
\psfrag{r4LL}{\small \hspace{-0.5cm} $R_4^g$, $R_4^f$}
\psfrag{r4LR}{\small \hspace{-0.5cm} $R_4^g$, $R_4^f$}
\psfrag{r5LL}{\small \hspace{-0.5cm} $R_5^g$, $R_5^f$}
\psfrag{r5LR}{\small \hspace{-0.5cm} $R_5^g$, $R_5^f$}
\includegraphics[width=15cm]{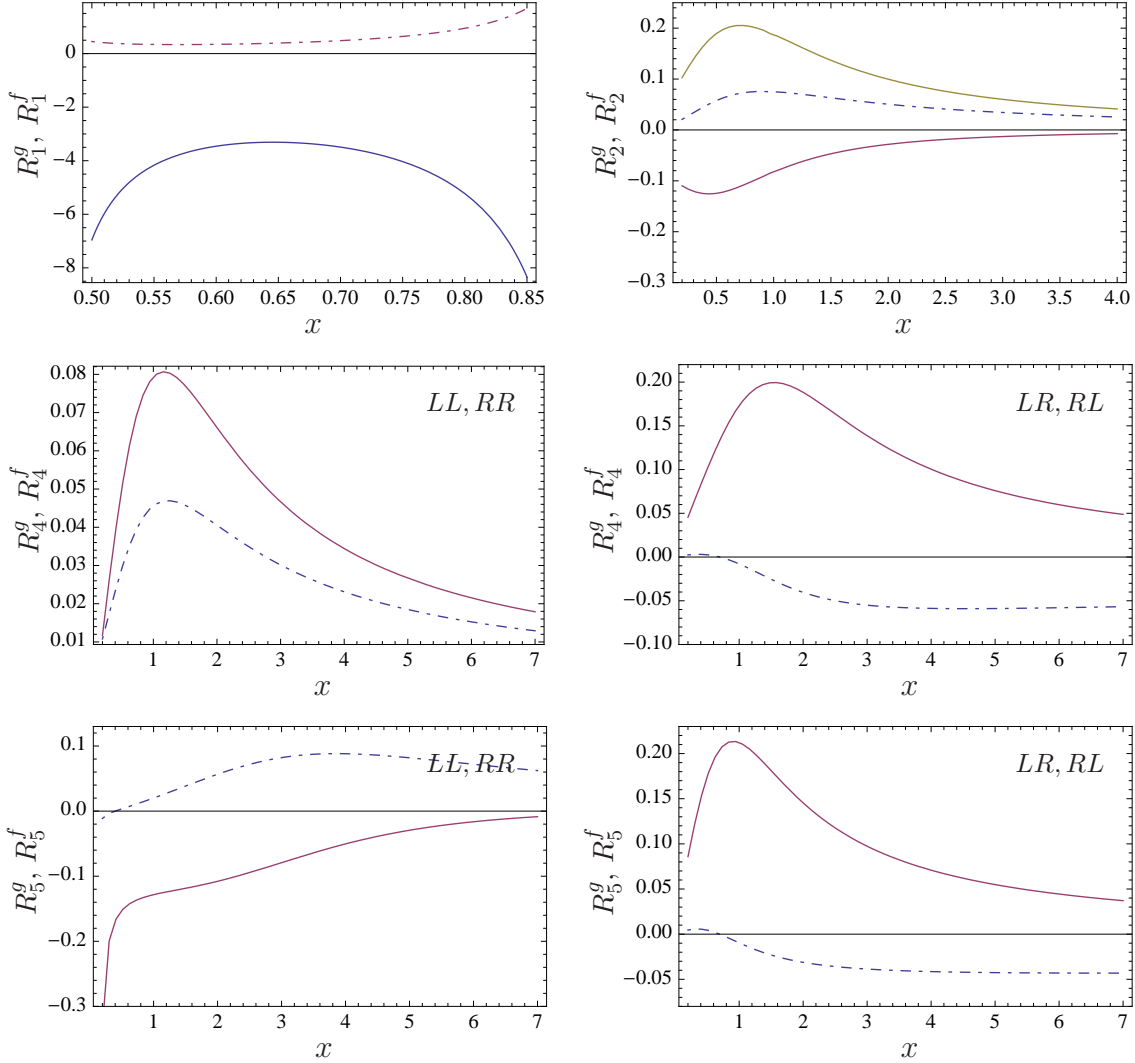}
\Text(-270,245)[lb]{\footnotesize $LL,RR$}
\Text(-47,245)[lb]{\footnotesize $LR,RL$}
\Text(-270,108)[lb]{\footnotesize $LL,RR$}
\Text(-47,108)[lb]{\footnotesize $LR,RL$}
\end{center}
\vspace{-0.5cm}
\caption{\small Relative size of the $m_t$-dependent contributions in the Mass Insertion Approximation. The ratios $R^g_i$ are shown as solid lines and $R^f_i$ as dashed lines. In the case of $R_2^g$, the upper solid curve shows the relative size of the contribution multiplying $\delta^{tt}_{RL}$ and the lower solid line the one multiplying $\delta^{tt}_{LR}$. In the case of coefficients $C_4$ and $C_5$, the two possibilities ($LL+RR$ or $RL+LR$ insertions) are displayed.}
\label{sizeMIA}
\end{figure}

In Fig.~\ref{sizeMIA} we show the Ratios $R^g_i$ and $R^f_i$ as a function of $x$ for $i=1,..,5$. We see that the only contribution that can partially overcome the suppression is the contribution to $C_1$. In the region $x \sim [0.5 - 1]$, this contribution could be up to $\sim 20\%$ of the whole NLO contribution, if $\mg\sim 1\ {\rm TeV}$ and $\delta^{tt}_{LR}\sim 0.2$. The same is obviously true for the coefficient $\widetilde C_1$. The rest of the contributions can be safely neglected over the whole parameter space in the degenerate scenario.

\subsection{Alignment}

A second scenario to consider is the ``aligned'' scenario, in which the squark mass eigenbasis is approximately aligned with the super-CKM basis. In this case, the diagonal squark masses can be different, with wide mass splittings. However, since the mixing angles are small (the rotation matrices are close to the identity), the off-diagonal entries (mass insertions) of the squark mass matrix in the super-CKM basis are small. An expansion in these mass insertions can be called the Non-degenerate MIA, or NDMIA.

From the full results of the Wilson coefficients presented in Appendix~\ref{sec:res} and in Ref.~\cite{JV}, the expressions of the WC's in the NDMIA can be obtained (see for example Section 6 of Ref.~\cite{JV}). These coefficients have the same structure as the MIA WC's, but they depend on a set of different squark masses instead of on a single variable $x$:
\eq{C_i^{\rm NDMIA} = \frac{\alpha_s^3}{12^3\pi \mg^2}
\Bigg[a_i({\bf x})+\frac{m_t}{\mg}\ g_i({\bf y})\ {\rm Re}(\delta^{tt}_{LR})
+\frac{m_t^2}{\mg^2}\ f_i({\bf y})+\cdots \Bigg] \bar\delta_{AB}^{sb}\bar\delta_{CD}^{sb}\ ,}
where the $m_t=0$ part depends on all 12 squark masses (as they all appear in the squark-quark loop): ${\bf x}=(x_{\tilde q_L},x_{\tilde q_R})$, and the $m_t\ne 0$ part depends on at most 6 masses: ${\bf y}=(x_{\tilde s_A},x_{\tilde b_B},x_{\tilde s_C},x_{\tilde b_D},x_{\tilde t_{L,R}})$. Also, the barred mass insertions $\bar\delta^{ij}_{XY}$ are the mass insertions normalized to the gluino mass (instead of to an average squark mass, which is now ill defined).

The large number of free parameters in this scenario does not allow for a consistent scan of the full parameter space. I any case, we are interested in finding ``tipical'' regions in which the new contributions are not negligible, so a full scan is not necessary. Since we are focusing on $B_s$ mixing, the most relevant squark masses are $m_{\tilde s}$ and $m_{\tilde b}$, so the strategy is the following. We scan randomly over squark masses (up to several times the gluino mass) for all squark masses except $m_{\tilde s}$ and $m_{\tilde b}$, and we plot in the $m_{\tilde s}$-$m_{\tilde b}$ plane the ratios $R_i^g=g_i/a_i$ and $R_i^f=f_i/a_i$.

\begin{figure}
\begin{center}
\psfrag{xsL}{\small $x_{\tilde s_L}$} \psfrag{xbL}{\small $x_{\tilde b_L}$}
\includegraphics[width=7cm]{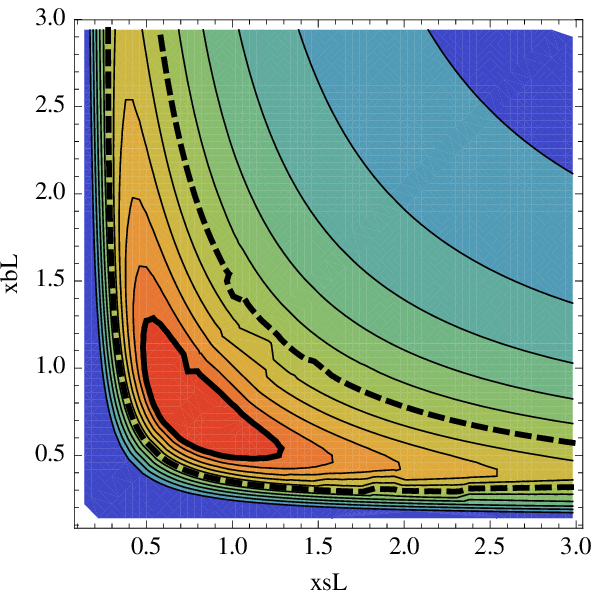}\hspace{1cm}
\includegraphics[width=7cm]{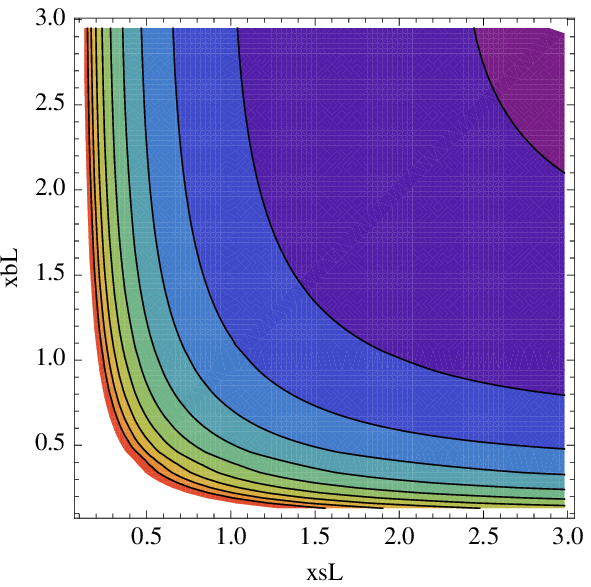}
\Text(-340,40)[lb]{\colorbox{gris}{\footnotesize 3}}
\Text(-300,55)[lb]{\colorbox{gris}{\footnotesize 2}}
\Text(-265,150)[lb]{\colorbox{gris}{\footnotesize 1}}
\Text(-35,150)[lb]{\colorbox{gris}{\footnotesize 0.5}}
\Text(-100,55)[lb]{\colorbox{gris}{\footnotesize 0.6}}
\Text(-120,40)[lb]{\colorbox{gris}{\footnotesize 0.7}}
\end{center}
\vspace{-0.5cm}
\caption{\small Relative size of the $m_t$-dependent contributions in the Non-degenerate Mass Insertion Approximation. Left: $R_1^g$ as a function of $x_{\tilde s_L}$ and $x_{\tilde b_L}$. The dashed line bounds the region $R_1^g>2$ and the thick solid line bounds the region $R_1^g>3$. Right: $R_5^g$ in the case of $LR,RL$ mixing. In both cases we take the rest of the squark masses $\sim {\cal O}(m_{\tilde g})$, as indicated in the text.}
\label{sizeNDMIA}
\end{figure}

In the left plot of Fig.~\ref{sizeNDMIA} we show the ratio $R_1^g$ as a function of  $m_{\tilde s_L}$ and $m_{\tilde b_L}$ for a random set of squark masses: $x_{\tilde u_L}=1.2$, $x_{\tilde u_R}=0.6$, $x_{\tilde d_L}=0.7$, $x_{\tilde d_R}=1.8$, $x_{\tilde s_R}=0.2$, $x_{\tilde c_L}=1.9$, $x_{\tilde c_R}=0.8$, $x_{\tilde b_R}=0.4$, $x_{\tilde t_L}=0.3$ $x_{\tilde t_R}=0.5$. The dashed line bounds the region where $R_1^g>2$ and the thick solid line indicates the region where $R_1^g>3$. In this region the size of the new contribution can be as high as $\sim 10$-$15\%$ of the whole NLO contribution to the Wilson coefficient $C_1$. The same is true again for $\widetilde C_1$. The ratio $R_1^f$, on the contrary, is in general small and can be neglected.

The contributions to the other Wilson coefficients are all negligible, or at least we have not found any generic region where the ratios $R_i$ for $i \ge 2$ are bigger than 1. As an example we show, in the right panel of Fig.~\ref{sizeNDMIA}, the ratio $R_5^g$ in the case of $LR,RL$ mixing. The squark masses used are the same as in the previous case. We see that this ratio is tipically smaller than 1, and around $\sim 0.5$ for $m_{\tilde s}, m_{\tilde b} \sim {\cal O}(m_{\tilde g})$. Very similar results are obtained for the rest of the coefficients.

\subsection{Hierarchical squark masses}

A third scenario to consider is the case of hierarchical squark masses. In this scenario, the first two generations of squarks are assumed to be much heavier than the rest of the supersymmetric spectrum, lying near the TeV scale.  Besides satisfying naturalness criteria \cite{hep-ph/9607394}, this scenario avoids flavor bounds from processes involving the first two families. This is due to the fact that contributions mediated by first two generation squarks are suppressed by a small ratio of masses, and the processes must proceed through the third generation, at the cost of containing an extra mass insertion.

A convenient parameterization of the squark rotation matrices in this scenario is the following \cite{arXiv:0812.3610}:

\eq{
\Gamma = \left(
\begin{array}{ccc|ccc}
1 & 0 & -\hat \delta^{13}_{LL} & 0 & 0 & 0\\[1.5mm]
0 & 1 & -\hat \delta^{23}_{LL} & 0 & 0 & 0\\[1.5mm]
\hat \delta^{13*}_{LL} \cos{\theta} & \hat \delta^{23*}_{LL} \cos{\theta} & \cos{\theta}
&\ \hat \delta^{13*}_{RR} \sin{\theta} e^{i\phi} & \hat \delta^{23*}_{RR} \sin{\theta} e^{i\phi} & \sin{\theta} e^{i\phi}\\[1.5mm]
\hline
&&&&&\\[-5mm]
0 & 0 & 0 & 1 & 0 & - \hat \delta^{13}_{RR}\\[1.5mm]
0 & 0 & 0 & 0 & 1 & - \hat \delta^{23}_{RR}\\[1.5mm]
 -\hat \delta^{13*}_{LL} \sin{\theta} e^{-i\phi} & -\hat \delta^{23*}_{LL} \sin{\theta} e^{-i\phi} & -\sin{\theta} e^{-i\phi}
 &  \hat \delta^{13*}_{RR} \cos{\theta} & \hat \delta^{23*}_{RR} \cos{\theta} & \cos{\theta} 
\end{array}
\right)
}\\
where it should be understood that there are two rotation matrices ($\Gamma_U$ and $\Gamma_D$), each with its own set of parameters. The $\delta's$ are dimensionless mass insertions, and $\theta_{U,D}$ and $\phi_{U,D}$ are mixing angles and phases that determine the diagonalization of the light-squark sector. The parameters involved in this scenario are: a common squark mass for the heavy sector $\ms_h\sim 10-100\ {\rm TeV}$, a common squark mass for the light sector $\ms_\ell\sim 1\ {\rm TeV}$, two mixing angles $\theta_{U,D}$ and two phases $\phi_{U,D}$ describing the mixing in the light sector, and the mass insertions $\hat\delta^{db}_{LL,RR}$, $\hat\delta^{sb}_{LL,RR}$, $\hat\delta^{ut}_{LL,RR}$ and $\hat\delta^{ct}_{LL,RR}$.

In this particular scenario, with this parameterization, and with no mass splitting between third generation $u$-type squarks, the corrections proportional to $m_t/\mg$ are zero:
\eq{\sum_k\, g(x_k)\, \Gamma_L^{kt*} \Gamma_R^{kt} = g(x_{\tilde t_L}) \sin{\theta} \cos{\theta} e^{i\phi}
-g(x_{\tilde t_R}) \sin{\theta} \cos{\theta} e^{i\phi}=0\ .}
The loop functions $f_i$ related to the corrections proportional to $m_t^2/\mg^2$ are, as happens in the other scenarios discussed previously, small with respect to their $m_t\to 0$ counterparts. We have checked that the ratios $R^f_i$ in this case are below $10\%$ in all cases.

\begin{figure}
\begin{center}
\psfrag{xtR}{\small $x_{\tilde t_R}$} \psfrag{xl}{\small $x_\ell$}
\includegraphics[width=4.7cm]{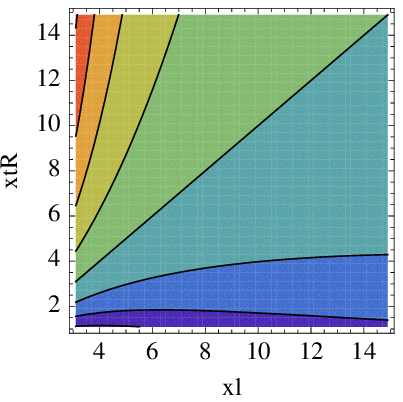}\hspace{0.5cm}
\includegraphics[width=4.7cm]{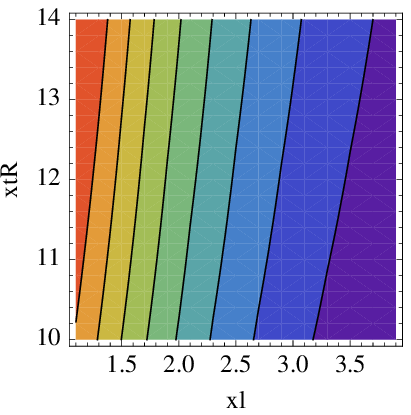}\hspace{0.5cm}
\includegraphics[width=4.7cm]{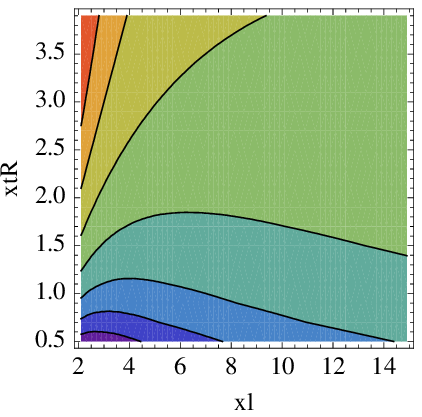}
\Text(-340,42)[lb]{\colorbox{gris}{\footnotesize -1}}
\Text(-360,85)[lb]{\colorbox{gris}{\footnotesize 0}}
\Text(-390,105)[lb]{\colorbox{gris}{\footnotesize 1}}
\Text(-330,26)[lb]{\colorbox{gris}{\tiny -2}}
\Text(-405,115)[lb]{\colorbox{gris}{\tiny 2}}
\Text(-180,40)[lb]{\colorbox{gris}{\footnotesize 3}}
\Text(-215,60)[lb]{\colorbox{gris}{\footnotesize 5}}
\Text(-235,80)[lb]{\colorbox{gris}{\footnotesize 7}}
\Text(-250,100)[lb]{\colorbox{gris}{\footnotesize 9}}
\Text(-100,110)[lb]{\colorbox{gris}{\footnotesize 0}}
\Text(-83,100)[lb]{\colorbox{gris}{\footnotesize -1}}
\Text(-53,55)[lb]{\colorbox{gris}{\footnotesize -2}}
\Text(-62,32)[lb]{\colorbox{gris}{\footnotesize -3}}
\Text(-85,27)[lb]{\colorbox{gris}{\tiny -4}}
\end{center}
\vspace{-0.5cm}
\caption{\small Relative size of the $m_t$-dependent contributions to $C_4$ in the Hierarchical squark mass scenario: $R_4^g$ as a function of $x_\ell$ and  $x_{\tilde t_R}$. We show an overview plot of a large region (left), and two particular regions where the ratio is large and positive (center) and large and negative (right). In these plots we have taken $x_h=1000$, $\theta=45^\circ$ and $\phi=0$.}
\label{sizeH}
\end{figure}

A mass splitting between $\tilde t_L$ and $\tilde t_R$ can change the situation. We choose the ``right handed'' stop to be the one whose mass deviates from $\tilde m_\ell$. In this case the ratios $R_i^g$ are nonzero for $C_1$, $C_4$ and $C_5$. We find that these ratios are not more than a few percent for $C_1$ and $C_5$, but they can be moderately large in the case of $C_4$. In Fig.~\ref{sizeH} we show the values of $R_4^g$ as a function of  $x_\ell$ and  $x_{\tilde t_R}$, taking for illustration for the other parameters the following values:  $x_h=1000$, $\theta=45^\circ$ and $\phi=0$. While this ratio is zero for $\tilde m_\ell = m_{\tilde t_R}$, it increases with the mass splitting, ranging from $\sim -4$ to $\sim 10$ in a relatively wide region.

\section{Conclusions}
\label{sec:con}

In this paper we have computed the top mass dependent two-loop squark-gluino corrections to the $\Delta F=2$ matching conditions. The expanded results for the Wilson coefficients up to order $m_t^2/\mg^2$ are given explicitly  in Appendix \ref{sec:res} (for the general case), and in Appendix \ref{sec:resMIA} in the Mass Insertion Approximation. This complements the results of Ref.~\cite{JV}, that were obtained in the limit $m_t/\mg\to 0$.

We have studied the relative size of this corrections compared to the $m_t\to 0$ results. For that matter we have considered three different SUSY scenarios. In the case of degeneracy, we find that the corrections are small and can be neglected in general, but there is a region for $\mg^2/M_s^2\sim 0.5 - 0.9$ where the contribution to the coefficient $C_1$ could be notable.

In the case of alignment, the situation is very similar. All contributions are negligible, but there is again a region in which the contribution to $C_1$ could partially overcome the suppression, corresponding to values of the squark masses $x_{\tilde s_L},x_{\tilde b_L}\sim 0.5-1.5$. This is true for a chosen set of values for the rest of the squark masses, which are specified in the text.

In the case of the hierarchical squark mass scenario, the situation is quite different. The correction linear in $m_t/\mg$ is strictly zero for degenerate stop quarks, given the chosen parameterization of the squark rotation matrices. When we allow for a mass splitting between left and right handed stops, these contributions are nonzero. We find that in this case they are negligible for all the coefficients except for $C_4$. The contribution to $C_4$ can become relevant in some regions of the parameter space specified in Fig.~\ref{sizeH}.

\subsubsection*{Acknowledgements}

J.V. is supported in part by ICREA-Academia funds.

\appendix

\section{Wilson Coefficients}
\label{sec:res}

In this section we give the explicit expressions for the Wilson coefficients. We omit the results for $\tilde C_{1,2,3}$, which are obtained from $C_{1,2,3}$ by exchanging $L \leftrightarrow R$ in the squark rotation matrices. Here we give only the $m_t$ dependent part of the Wilson coefficients:
\eq{C_i^{m_t}=C_i-C_i|_{m_t=0}}
The functions depend on the squark masses through the squared mass ratios $x_i\,$, defined as $x_i\equiv \tilde m_i^2/m_{\tilde g}^2$. A sum over $i,j,k$ is understood, running over all right and left-handed squarks of $u$ or $d$ type depending on the case. We also define the product of squark rotation matrices:
\eq{\Gamma^{ijk}_{ABCDEF} \equiv \Gamma_A^{is*} \Gamma_B^{ib}\, \Gamma_C^{js*} \Gamma_D^{jb}\, \Gamma_E^{kt*} \Gamma_F^{kt}}
 Expanding the Wilson coefficients in powers of $m_t/m_{\tilde g}$ we have, up to order $(m_t/m_{\tilde g})^2$:

\eqa{
&&C_1^{m_t} =\frac{\alpha_s^3}{12^3\pi m_{\tilde g}^2} \Bigg[
\frac{m_t}{m_{\tilde g}} g_1^{ijk}\,(\Gamma^{ijk}_{LLLLLR}+\Gamma^{ijk}_{LLLLRL})
+\frac{m_t^2}{m_{\tilde g}^2} f_1^{ijk}\,(\Gamma^{ijk}_{LLLLLL}+\Gamma^{ijk}_{LLLLRR})
+\cdots\Bigg]\nn\\[2mm]
&&C_2^{m_t} = \frac{\alpha_s^3}{12^3\pi m_{\tilde g}^2} \Bigg[
\frac{m_t}{m_{\tilde g}} (g_2^{ijk}\,\Gamma^{ijk}_{RLRLLR}+h_2^{ijk}\,\Gamma^{ijk}_{RLRLRL})\nn\\
&&\hspace{3.5cm}+\frac{m_t^2}{m_{\tilde g}^2} f_2^{ijk}\,(\Gamma^{ijk}_{RLRLLL}+\Gamma^{ijk}_{RLRLRR})
+\cdots \Bigg]\hspace{1cm}\\[2mm]
&&C_4^{m_t} = \frac{\alpha_s^3}{12^3\pi m_{\tilde g}^2} \Bigg[
\frac{m_t}{m_{\tilde g}} \bigg( g_{41}^{ijk}\,(\Gamma^{ijk}_{LLRRLR}+\Gamma^{ijk}_{LLRRRL})
+ g_{42}^{ijk}\,(\Gamma^{ijk}_{LRRLLR}+\Gamma^{ijk}_{LRRLRL})\bigg)\nn\\
&&\hspace{2.6cm}+\frac{m_t^2}{m_{\tilde g}^2} \bigg(
f_{41}^{ijk}\,\,(\Gamma^{ijk}_{LLRRLL}+\Gamma^{ijk}_{LLRRRR})
+f_{42}^{ijk}\,\,(\Gamma^{ijk}_{LRRLLL}+\Gamma^{ijk}_{LRRLRR})\bigg)
+\cdots\Bigg]\nn\\[2mm]
&&C_5^{m_t} = \frac{\alpha_s^3}{12^3\pi m_{\tilde g}^2} \Bigg[
\frac{m_t}{m_{\tilde g}} \bigg( g_{51}^{ijk}\,(\Gamma^{ijk}_{LLRRLR}+\Gamma^{ijk}_{LLRRRL})
+ g_{52}^{ijk}\,(\Gamma^{ijk}_{LRRLLR}+\Gamma^{ijk}_{LRRLRL})\bigg)\nn\\
&&\hspace{2.6cm}+\frac{m_t^2}{m_{\tilde g}^2} \bigg(
f_{51}^{ijk}\,\,(\Gamma^{ijk}_{LLRRLL}+\Gamma^{ijk}_{LLRRRR})
+f_{52}^{ijk}\,\,(\Gamma^{ijk}_{LRRLLL}+\Gamma^{ijk}_{LRRLRR})\bigg)
+\cdots\Bigg]\nn
}
and $C_3^{m_t} = -3/17\,C_2^{m_t}$. The loop functions can be written as:

\eq{
\begin{array}{lll}
\ds f_1^{ijk} = \frac{F_1^{ik}-F_1^{jk}}{x_i-x_j}\ ; \hspace{1cm}
& \ds g_1^{ijk} = \frac{G_1^{ik}-G_1^{jk}}{x_i-x_j}\ ; \hspace{1cm}
&\ds f_2^{ijk} = \frac{F_2^{ik}-F_2^{jk}}{x_i-x_j}\\[5mm]
\ds g_2^{ijk} = \frac{G_2^{ik}-G_2^{jk}}{x_i-x_j}\ ; \hspace{1cm}
& \ds h_2^{ijk} = \frac{H_2^{ik}-H_2^{jk}}{x_i-x_j}\ ; \hspace{1cm}
&\ds f_{41}^{ijk} = \frac{F_{41}^{ik}-F_{41}^{jk}}{x_i-x_j}\\[5mm]
\ds g_{41}^{ijk} = \frac{G_{41}^{ik}-G_{41}^{jk}}{x_i-x_j}\ ; \hspace{1cm}
&\ds f_{42}^{ijk} = \frac{F_{42}^{ik}-F_{42}^{jk}}{x_i-x_j}\ ; \hspace{1cm}
& \ds g_{42}^{ijk} = \frac{G_{42}^{ik}-G_{42}^{jk}}{x_i-x_j}\\[5mm]
\ds f_{51}^{ijk} = \frac{F_{51}^{ik}-F_{51}^{jk}}{x_i-x_j}\ ; \hspace{1cm}
& \ds g_{51}^{ijk} = \frac{G_{51}^{ik}-G_{51}^{jk}}{x_i-x_j}\ ; \hspace{1cm}
&\ds f_{52}^{ijk} = \frac{F_{52}^{ik}-F_{52}^{jk}}{x_i-x_j}\\[5mm]
\ds g_{52}^{ijk} = \frac{G_{52}^{ik}-G_{52}^{jk}}{x_i-x_j}
\end{array}
}
All these can be expressed in terms of 5 master functions, that we have chosen to be $F_1$, $F_2$, $G_1$, $G_2$ and $A$, such that:

\eq{
\begin{array}{lll}
\ds H_2^{ik} = 17 G_1^{ik}-\frac{13}{2} G_2^{ik}+\frac{187}{2} A^{ik}\ ; \hspace{0.3cm}
& \ds F_{41}^{ik} = -\frac{12}{11} F_1^{ik} + \frac{486}{187} F_2^{ik}\ ; \hspace{0.3cm}
&\ds F_{42}^{ik} = -2 F_1^{ik} + \frac{4}{17} F_2^{ik}\\[5mm]
\ds G_{41}^{ik} = 21 G_1^{ik}-\frac{243}{34} G_2^{ik}+\frac{243}{2} A^{ik}\ ; \hspace{0.3cm}
& \ds G_{42}^{ik} = 11 G^{ik} - \frac{11}{17} G_2^{ik}\ ; \hspace{0.3cm}
&\ds F_{51}^{ik} = \frac{20}{11} F_1^{ik} - \frac{18}{187} F_2^{ik}\\[5mm]
\ds G_{51}^{ik} = G_1^{ij} + \frac{9}{34} G_2^{ik} - \frac{9}{2} A^{ik}\ ; \hspace{0.3cm}
&\ds F_{52}^{ik} = \frac{15}{11} F_{42}^{ik}\ ; \hspace{0.3cm}
& \ds G_{52}^{ik} = \frac{15}{11} G_{42}^{ik}
\end{array}
}
The master functions are given by:

\eqa{
A^{ik} &=& -24\Bigg[ \frac{2x_i}{(x_i-1)^3}\log{x_i} + \frac{(3x_k-1)x_i^2 - 3x_k^2 x_i - x_i +xk^2+x_k}{(x_i-1)^2(x_k-1)^2}\Bigg]\\[5mm]
F_1^{ik} &=& 12\Bigg[ \frac{2 x_i x_k (19 x_i^2-57 x_i+68)}{(x_i-1)^3}\ \Li(1-x_k) + \frac{2x_k(11 x_i+19)}{(x_i-1)^3}\ \Li(\textstyle 1-\frac{x_i}{x_k} )\nn\\
&&\hspace{-0.7cm} - \frac{(11x_k+19)x_i^3-(33x_k+57)x_i^2-(11x_k^3-33x_k-68)x_i}{(x_i-1)^3(x_k-1)^3}\ x_k\log^2{x_k}\\
&&\hspace{-0.7cm} + \frac{19x_k^2 - 57x_k + 68}{(x_i-1)^3(x_k-1)^3} x_k^2 \log^2{x_k} - \frac{2x_k((19 x_k-15)x_i^2-(49x_k-26)x_i+19)}{(x_i-1)^2(x_k-1)^2} \log{x_k}\nn\\
&&\hspace{-0.7cm} - \frac{2x_i(11x_i+19)}{(x_i-1)^3}\log{x_i} - 2\frac{(11x_k+4)x_i^2-(11x_k^2-15x_k+34)x_i- x_k(19x_k-34)}{(x_i-1)^2(x_k-1)^2}\Bigg] \nn\\[5mm]
F_2^{ik} &=& 816\Bigg[ \frac{2 x_i x_k (x_i^2-3 x_i+3)}{(x_i-1)^3}\ \Li(1-x_k) + \frac{2x_k}{(x_i-1)^3}\ \Li(\textstyle 1-\frac{x_i}{x_k} )\nn\\
&&\hspace{-0.7cm} - \frac{x_i^3-3x_i^2+3x_i-x_k^3+3x_k^2-3 x_k}{(x_i-1)^3(x_k-1)^3}\ x_k\log^2{x_k}
- \frac{x_k(x_i-2)(2x_kx_i-x_i-1)}{(x_i-1)^2(x_k-1)^2} \log{x_k}\nn\\
&&\hspace{-0.7cm} - \frac{2x_i}{(x_i-1)^3}\log{x_i} - \frac{x_i^2+(x_k-3)x_i-2x_k^2 +3 x_k}{(x_i-1)^2(x_k-1)^2}\Bigg]\\[5mm]
G_1^{ik} &=& 24\Bigg[ \frac{2 x_i (-2x_kx_i^2+(6x_k+13)x_i-19x_k+2)}{(x_i-1)^3}\ \Li(1-x_k)\nn\\ 
&&\hspace{-0.7cm} + \frac{2(13x_i+2)(x_i-x_k)}{(x_i-1)^3}\ \Li(1-{\textstyle \frac{x_i}{x_k}}) - \frac{2x_k^3-6x_k^2+45x_k+4}{(x_i-1)^3(x_k-1)^3}\ x_k\log^2{x_k}\nn\\
&&\hspace{-0.7cm} + \frac{3(13x_k+2)x_kx_i+13x_k^3-156 x_k^2+21x_k-13}{(x_i-1)^3(x_k-1)^3} x_i^2 \log^2{x_k} - \frac{x_i(13x_i+2)}{(x_i-1)^3}\log^2{x_i}\nn\\
&&\hspace{-0.7cm} - \frac{13x_k^4 - 41x_k^3 - 72x_k^2 - 37x_k + 2}{(x_i-1)^3(x_k-1)^3} x_i \log^2{x_k}	 - \frac{2x_i(13x_i+2)}{(x_i-1)^3} \log{x_i}\log{x_k}\nn\\
&&\hspace{-0.7cm} - \frac{(2x_k^3-6x_k^2+110x_k-5)x_i-2x_k^3-9x_k^2-80x_k-10}{(x_i-1)(x_k-1)^3} x_k\log{x_k} + \frac{x_i(89x_i+12)}{(x_i-1)^3}\log{x_i} \nn\\
&&\hspace{-0.7cm}  + \frac{(219x_k-17)x_i^2-(219x_k^2+185)x_i + x_k(17x_k+185)}{2(x_i-1)^2(x_k-1)^2}\Bigg] \\[5mm]
G_2^{ik} &=& 816\Bigg[ \frac{2 x_i (x_i-x_k)}{(x_i-1)^3}\ \Li(1-x_k) + \frac{2x_i(x_i-x_k)}{(x_i-1)^3}\ \Li(\textstyle 1-\frac{x_i}{x_k} )\nn\\
&&\hspace{-0.7cm} + \frac{3x_k^2x_i^3+(x_k^3-12x_k^2+3x_k-1)x_i^2-(x_k^4-3x_k^3-6x_k^2-x_k)x_i-3x_k^2}{(x_i-1)^3(x_k-1)^3}\ \log^2{x_k}\nn\\
&&\hspace{-0.7cm} - \frac{2x_i^2}{(x_i-1)^3}\ \log{x_i}\log{x_k}
- \frac{(7x_k-1)x_i-x_k^2-5x_k}{(x_i-1)(x_k-1)^3}\ x_k\log{x_k}
- \frac{x_i^2}{(x_i-1)^3}\log^2{x_i}\nn\\
&&\hspace{-0.7cm} + \frac{6x_i^2}{(x_i-1)^3}\ \log{x_i}
+ \frac{(7x_k-1)x_i^2-(7x_k^2+5)x_i + x_k(x_k+5)}{(x_i-1)^2(x_k-1)^2}\Bigg]
}
Here, the function $\Li(x)$ denotes the dilogarithm defined in the usual way:
\eq{\Li(x)=-\int_0^x dt\ \frac{\log(1-t)}{t}\ .}

\section{Wilson Coefficients in the MIA}
\label{sec:resMIA}

In this Section we present the expressions for the Wilson coefficients in the Mass Insertion Approximation. Again, we present only the $m_t$ dependent part of the Wilson coefficients:
\eq{C_i^{m_t}=C_i-C_i|_{m_t=0}}
The coefficients $\tilde C_{1,2,3}$ are obtained from $C_{1,2,3}$ by exchanging $L \leftrightarrow R$ in the squark rotation matrices, and $C_3^{m_t} = -3/17\,C_2^{m_t}$. The variable $M_s^2$ denotes the squared squark mass, meaning any of the common diagonal entries in the squark mass matrix in the super CKM basis (the diagonal masses are corrected by the mass insertions). The coefficient functions depend on the variable $x=m_{\tilde g}^2/M_s^2$. The rest of the Wilson coefficients are given by:
\eqa{
C_1^{m_t} &=& \frac{\alpha_s^3}{12^3\pi M_s^2} \frac{m_t}{m_{\tilde g}} \frac1{3(x-1)^6}
\Bigg[
\bigg(11360 x + 2056 x^2 - 14496 x^3 + 1144 x^4 - 64 x^5\nn\\
&-& 3744 x \log{x} - 15696 x^2 \log{x} - 14976 x \Li{(1 - x)} - 28224 x^2 \Li{(1 - x)}\bigg) {\rm Re}(\delta^{tt}_{LR})\nn\\
&+&\frac{m_t}{m_{\tilde g}}\bigg(-10208 x - 44032 x^2 + 51576 x^3 + 2816 x^4 - 152 x^5- 2592 x^2 \log{x}\nn\\
&-& 21168 x^3 \log{x} + 6336 x \Li{(1 - x)} + 36864 x^2 \Li{(1 - x)}\bigg)
\Bigg] (\delta_{LL}^{sb})^2\\
C_2^{m_t}&=& \frac{\alpha_s^3}{12^3\pi M_s^2} \frac{m_t}{m_{\tilde g}} \frac1{3(x-1)^6}
\Bigg[
\bigg(15232 x - 2856 x^2 - 13056 x^3 + 680 x^4 - 4896 x \log{x}\nn\\
&-& 17136 x^2 \log{x} - 19584 x \Li{(1-x)} - 29376 x^2 \Li{(1-x)}
\bigg) \delta^{tt}_{LR}
+ \bigg(-2448 x \nn\\
&+& 36040 x^2 -38352 x^3 + 5304 x^4 - 544 x^5 - 22032 x^2 \log{x} - 48960 x^2 \Li{(1-x)}
\bigg) \delta^{tt\,*}_{LR}\nn\\
&+&\frac{m_t}{m_{\tilde g}}\bigg(
-143888 x^2 + 135456 x^3 + 8976 x^4 - 544 x^5 + 4896 x^2 \log{x} - 58752 x^3 \log{x}\nn\\
&+& 97920 x^2 \Li{(1-x)}\bigg)
\Bigg] (\delta_{RL}^{sb})^2\\
C_4^{m_t} &=& \frac{\alpha_s^3}{12^3\pi M_s^2} \frac{m_t}{m_{\tilde g}} \frac1{3(x-1)^6}
\Bigg\{
\Bigg[
\bigg(
20832 x + 84000 x^2 - 117792 x^3 + 14304 x^4 - 1344 x^5\nn\\
&-& 8640 x \log{x} - 84672 x^2 \log{x} - 34560 x \Li{(1-x)} - 172800 x^2 \Li{(1-x)}
\bigg) {\rm Re}(\delta^{tt}_{LR})\nn\\
&+&\frac{m_t}{m_{\tilde g}}\bigg(
11136 x - 325920 x^2 + 295776 x^3 + 20256 x^4 - 1248 x^5 + 15552 x^2 \log{x}\nn\\
&-& 129600 x^3 \log{x} - 6912 x \Li{(1-x)} + 214272 x^2 \Li{(1-x)}
\bigg)
\Bigg] (\delta_{LL}^{sb})(\delta_{RR}^{sb})\nn\\
&+&\Bigg[
\bigg(
-19712 x + 3696 x^2 + 16896 x^3 - 880 x^4 + 6336 x \log{x} + 22176 x^2 \log{x}\nn\\
&+& 25344 x \Li{(1-x)} + 38016 x^2 \Li{(1-x)}
\bigg) {\rm Re}(\delta^{tt}_{LR})
+\frac{m_t}{m_{\tilde g}}\bigg(
20416 x + 54208 x^2 \nn\\
&-& 71280 x^3 - 3520 x^4 + 176 x^5 + 6336 x^2 \log{x} + 28512 x^3 \log{x} - 12672 x \Li{(1-x)}\nn\\
&-& 50688 x^2 \Li{(1-x)}\bigg)\Bigg] (\delta_{LR}^{sb})(\delta_{RL}^{sb})
\Bigg\}\\
C_5^{m_t} &=& \frac{\alpha_s^3}{12^3\pi M_s^2} \frac{m_t}{m_{\tilde g}} \frac1{3(x-1)^6}
\Bigg\{
\Bigg[
\bigg(
19424 x + 544 x^2 - 21408 x^3 + 1504 x^4 - 64 x^5\nn\\
&-& 6336 x \log{x} - 24768 x^2 \log{x} - 25344 x \Li{(1-x)} - 43776 x^2 \Li{(1-x)}
\bigg) {\rm Re}(\delta^{tt}_{LR})\nn\\
&+&\frac{m_t}{m_{\tilde g}}\bigg(
-18560 x - 66208 x^2 + 80736 x^3 + 4256 x^4 - 224 x^5 - 5184 x^2 \log{x}\nn\\
&-& 32832 x^3 \log{x} + 11520 x \Li{(1-x)} + 57600 x^2 \Li{(1-x)}
\bigg)
\Bigg] (\delta_{LL}^{sb})(\delta_{RR}^{sb})\nn\\
&+&\Bigg[
\bigg(
-26880 x + 5040 x^2 + 23040 x^3 - 1200 x^4 + 8640 x \log{x} + 30240 x^2 \log{x}\nn\\
&+& 34560 x \Li{(1-x)} + 51840 x^2 \Li{(1-x)}
\bigg) {\rm Re}(\delta^{tt}_{LR})
+\frac{m_t}{m_{\tilde g}}\bigg(
27840 x + 73920 x^2\nn\\
&-& 97200 x^3 - 4800 x^4 + 240 x^5 + 8640 x^2 \log{x} + 38880 x^3 \log{x} - 17280 x \Li{(1-x)}\nn\\
&-& 69120 x^2 \Li{(1-x)}
\bigg)\Bigg] (\delta_{LR}^{sb})(\delta_{RL}^{sb})
\Bigg\}
}



\end{document}